\documentclass[conference,compsoc,11pt]{IEEEtran}

\ifCLASSOPTIONcompsoc
  \usepackage[nocompress]{cite}
\else
  \usepackage{cite}
\fi
\ifCLASSINFOpdf

\else
 
\fi
\usepackage{setspace}
\usepackage[absolute,overlay]{textpos}
\usepackage{graphicx}

\begin{document}


\title{AI Governance and Accountability: An Analysis of Anthropic's Claude}

\author{\IEEEauthorblockN{Aman Priyanshu}
\IEEEauthorblockA{Privacy Engineering,\\
School of Computer Science,\\
Carnegie Mellon University \\
Email: apriyans@andrew.cmu.edu}
\and
\IEEEauthorblockN{Yash Maurya}
\IEEEauthorblockA{Privacy Engineering,\\
School of Computer Science,\\
Carnegie Mellon University \\
Email: ymaurya@andrew.cmu.edu}
\and
\IEEEauthorblockN{Zuofei Hong}
\IEEEauthorblockA{Privacy Engineering,\\
School of Computer Science,\\
Carnegie Mellon University \\
Email: zuofeih@andrew.cmu.edu}}


%


\maketitle

\begin{abstract}
As AI systems become increasingly prevalent and impactful, the need for effective AI governance and accountability measures is paramount. This paper examines the AI governance landscape, focusing on Anthropic's Claude, a foundational AI model. We analyze Claude through the lens of the NIST AI Risk Management Framework and the EU AI Act, identifying potential threats and proposing mitigation strategies. The paper highlights the importance of transparency, rigorous benchmarking, and comprehensive data handling processes in ensuring the responsible development and deployment of AI systems. We conclude by discussing the social impact of AI governance and the ethical considerations surrounding AI accountability.
\end{abstract}


%
\IEEEpeerreviewmaketitle

\section{Introduction}

Artificial Intelligence (AI) has become an integral part of modern society, pervading diverse domains from complex computational tasks and generating reports to mass communication, hiring decisions, and marketing efforts. As AI systems continue to grow in sophistication and influence, their impact expands across numerous spheres, shaping decision-making processes, information dissemination, and human interactions on an unprecedented scale. 
\begin{figure}[t]
    \centering
    \includegraphics[width=0.475\textwidth]{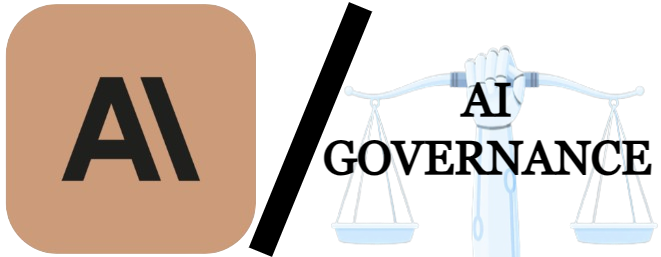}
    \caption{Anthropic's Claude is one of the most popular large language model chatbots available to the everyday consumer. This paper presents a study of its practices and conduct through the lens of AI governance.}
    \label{fig:enter-label}
\end{figure}
In this new era of AI, foundation models have assumed a significant role. Models, such as Anthropic's Claude, a large language model (LLM) capable of understanding and generating human-like text, exhibit unique potential for quick, effective, and scalable communication efforts. The customer reach of these LLMs, like Claude, has drastically increased over the years, and their influence on individuals' lives is expected to continue growing. Anthropic has announced several partnerships with prominent companies such as Scale\cite{Scale}, Zoom\cite{Zoom}, BCG\cite{BCG}, AWS\cite{Accenture}, Accenture\cite{Accenture}, SKT Telecom\cite{Sktelecom}, and Keif Studio\cite{Kief}, further amplifying the impact of their AI systems on people's lives, often without their knowledge of interacting with an AI system.

These LLMs are crucial as they underpin many AI systems, influencing outcomes and decision-making processes in areas that directly affect individuals and societies. Due to their unprecedented potential for impact, these foundation models must be evaluated for the risks and challenges they may pose to society.

These challenges motivate the need for AI governance - the processes, policies, and practices aimed at ensuring the responsible development, deployment, and use of AI systems. The importance of AI governance lies in its ability to ensure the responsible development and deployment of AI systems, safeguarding against potential harms and unintended consequences. Accountability is a key aspect of AI governance, as it helps establish trust and ensures that AI systems are designed and used in an ethical and transparent manner. Frameworks such as the NIST AI Risk Management Framework and the EU AI Act provide guidelines and standards for assessing, categorizing, and managing AI risks. These frameworks enable stakeholders to develop appropriate governance measures by offering structured approaches to identify, analyze, and mitigate potential threats associated with AI systems.

\begin{figure}
    \centering
    \includegraphics[width=0.475\textwidth]{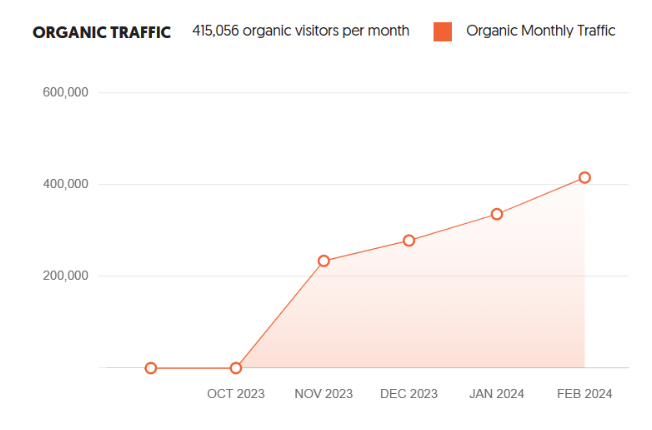}
    \caption{Rapidly growing customer visits on their Claude's web interface}
    \label{fig:enter-label}
\end{figure}

\begin{figure}
    \centering
    \includegraphics[width=0.3\textwidth]{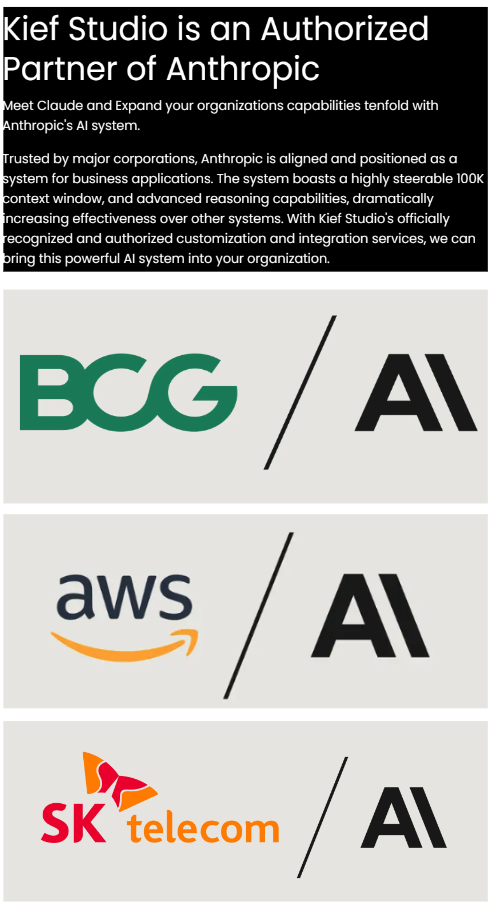}
    \caption{Some of Anthropic's Partnerships}
    \label{fig:enter-label}
\end{figure}

In this paper, we analyze Anthropic's Claude through the lens of these frameworks, identifying potential threats and proposing mitigation strategies. We also focus on their Constitutional AI paradigm. By examining Claude as not only an AI product but a foundational model, we aim to provide insights that can inform the broader AI governance discourse and contribute to the responsible advancement of AI technologies. The key objectives of this paper are:
\vspace{-1em}
\begin{enumerate}
\item To analyze Anthropic's Claude through established AI governance frameworks like:
\begin{itemize}
    \item NIST
    \item EU AI Act
\end{itemize}
\item To identify potential threats and risks posed by Claude
\item To propose mitigation strategies for these risks
\item To examine Anthropic's Constitutional AI paradigm
\item To provide insights for broader AI governance discourse
\end{enumerate}

\section{Organization of this Report}

This report is organized into several key sections to provide a comprehensive analysis of AI governance and accountability, with a focus on Anthropic's Claude model. The \textbf{introduction} sets the stage by highlighting the growing importance of AI governance as AI systems become increasingly prevalent and influential in various domains. It emphasizes the role of foundation models, such as Claude, in shaping decision-making processes and the need for effective governance measures to ensure responsible AI development and deployment.

The \textbf{literature review} section explores the current state of AI governance, discussing various frameworks and guidelines, such as the NIST AI Risk Management Framework and the EU AI Act. It also dives into recent literature on AI governance themes, knowledge gaps, and future agendas, as well as the challenges arising from the growing number of AI ethics documents produced by corporations, governments, and NGOs.

The \textbf{preliminaries} section provides a brief overview of key concepts, including artificial intelligence, large language models, and Anthropic's Claude. It also introduces Constitutional AI, a framework employed by Anthropic to align the model's outputs with predefined ethical principles and values.

The \textbf{threat analysis} section forms the main crux of the report, identifying and discussing potential threats and issues associated with Claude. This section focuses on specific risks, such as the lack of transparency in privacy policies, potential for hallucinations and biases in outputs, concerns about third-party data usage, and the implications of Constitutional AI. The analysis is conducted through the lens of the \textbf{NIST AI Risk Management Framework}, examining aspects of governance, risk mapping, and impact characterization. Additionally, the \textbf{EU AI Act} is used to categorize the identified risks based on their severity and potential consequences.

Building upon the threat analysis, the report proposes \textbf{mitigation strategies and resolution approaches} to address the identified risks. These strategies include enhancing transparency in privacy policies, establishing rigorous benchmarks for hallucination and bias, and developing comprehensive remediation processes for data deletion and model unlearning. The discussion section explores the broader implications of these mitigation strategies for the AI governance landscape and the social impact of AI systems.

The \textbf{conclusion} summarizes the key findings and highlights the importance of ongoing collaboration, adaptation, and learning in the evolution of AI governance. It emphasizes the need for aligning AI systems with ethical principles and societal values to foster public trust and support the responsible advancement of AI technologies.

Finally, the report acknowledges its \textbf{limitations and discusses ethical considerations} in the development and deployment of AI systems. It stresses the importance of prioritizing ethical principles throughout the AI lifecycle and engaging in ongoing research and stakeholder collaboration to address the ethical implications of AI and develop robust governance frameworks.

We organize this report in this manner, to provide a comprehensive and structured analysis of AI governance and accountability, focusing on the specific risks associated with Anthropic's Claude model (plus its Constitutional AI efforts) and propose actionable mitigation strategies to ensure responsible AI development and deployment.

\section{Literature Review}

AI governance has gained significant attention in recent years, with various frameworks and guidelines proposed to address the risks and challenges associated with AI systems. The NIST AI Risk Management Framework \cite{nist_ai_24} provides a comprehensive approach to identifying, assessing, and managing AI risks, emphasizing the importance of governance, risk mapping, and impact characterization. Similarly, the EU AI Act \cite{EUAI2024Apr} categorizes AI systems based on their risk levels, imposing specific requirements and obligations for high-risk systems.

Beyond regulations and risk frameworks, recent literature has further explored the themes, knowledge gaps, and future agendas in AI governance \cite{AIGovThemes2022}. Key themes identified include technology, stakeholders and context, regulation, and processes. However, knowledge gaps remain, such as limited understanding of AI governance implementation, lack of attention to context, uncertain effectiveness of ethical principles and regulation, and insufficient operationalization of processes \cite{10.1145/3375627.3375804}. The growing number of AI ethics documents produced by corporations, governments, and NGOs since 2016 raises important considerations \cite{10.1145/3375627.3375804, papagiannidis2023toward}. Challenges may arise from the relative homogeneity of the documents' creators, and the varied impacts and success factors of these documents on the AI governance landscape warrant examination \cite{papagiannidis2023toward}. Translating AI ethical principles into practicable governance processes is crucial, and a concise AI governance definition can help identify the constituent parts of this complex problem \cite{mantymaki2022defining}.

Previous works on AI auditing and self-governance have highlighted the need for transparency, accountability, and continuous monitoring of AI systems. For example, Raji et al. \cite{raji2020closing} propose a framework for closing the AI accountability gap, emphasizing the importance of external audits and stakeholder engagement. Additionally, the development of privacy regulations, such as the General Data Protection Regulation (GDPR) \cite{gdpr24, goddard2017eu}, has demonstrated the importance of proactive measures and the need for ongoing adaptation to address emerging risks.

As AI governance continues to evolve, it is essential to learn from the successes and challenges of privacy regulations and apply these lessons to the development of AI accountability measures. We study Claude, one of the most popular AI models, through the lens of prior literature and recommended frameworks, as it has the capacity for large-scale harm if not studied ethically \cite{adetayo2024microsoft, uppalapati2024comparative}.

\section{Preliminaries}

Artificial Intelligence (AI) refers to the development of computer systems that can perform tasks that typically require human intelligence, such as visual perception, speech recognition, decision-making, and language translation. AI encompasses various subfields, including machine learning, natural language processing, and computer vision.

Large Language Models (LLMs) are a type of AI model that have gained significant attention in recent years. LLMs are trained on vast amounts of text data, enabling them to generate human-like text, answer questions, and perform various language-related tasks. These models, such as OpenAI's GPT series \cite{brown2020language} and Google's BERT \cite{devlin2019bert}, have demonstrated remarkable capabilities and have been applied to a wide range of applications.

Anthropic's Claude is a foundational AI model that aims to push the boundaries of AI capabilities while prioritizing safety and ethical considerations. Claude is designed to be a multi-purpose AI assistant, capable of engaging in open-ended conversations, answering questions, and assisting with various tasks. One of the key features of Claude is its grounding in Constitutional AI \cite{ClaudeCons2024Apr, Bai2022Dec}, a framework that aims to ensure the model's outputs align with predefined ethical principles and values.

\subsection{Constitutional AI}

\begin{figure*}[h!]
  \includegraphics[width=\textwidth]{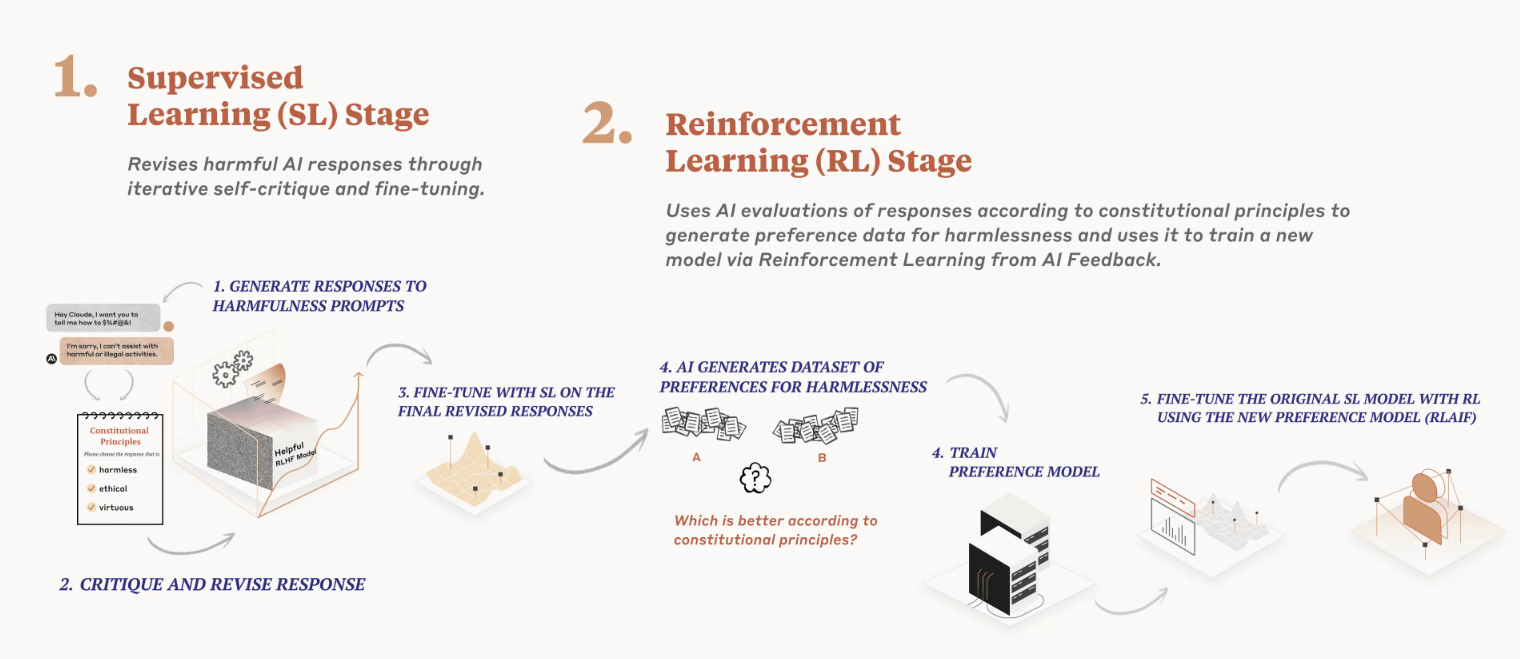}
  \caption{Anthropic's Constitutional AI training process}
  \label{fig:constitutional_ai}
\end{figure*}

Anthropic's Constitutional AI incorporates a set of principles, or a "constitution," to guide the model's behavior during the training process(Figure~\ref{fig:constitutional_ai})\cite{ClaudeCons2024Apr}. The constitution is used in two phases: first, the model is trained to critique and revise its own responses based on the principles; second, the model undergoes reinforcement learning using AI-generated feedback derived from the principles, rather than human feedback. This approach has been shown to produce models that are both more helpful and more harmless compared to models trained solely with human feedback\cite{Bai2022Dec}.

The principles in Anthropic's constitution are drawn from various sources, including the UN Declaration of Human Rights\cite{UnitedNations2024Apr}, trust and safety best practices inspired from Apple's Terms of Service, DeepMind's Sparrow principles\cite{DeepmindSparrow}, and values that encourage the consideration of Non-Western perspectives. These principles cover a wide range of topics, from the protection of human rights and the promotion of equality to the avoidance of harmful, deceptive, or offensive content.

In October 2023, Anthropic partnered with the Collective Intelligence Project to run a public input process involving approximately 1,000 Americans to draft a constitution for an AI system\cite{CollectiveConstitution}. The raw data from the survey is presented in the report by the Polis Center\cite{SurveyPolisReport}. The resulting publicly sourced constitution\cite{PubliclySourcedConstitution} had a moderate degree of overlap with Anthropic's in-house constitution (roughly 50\% overlap in concepts and values). However, the public constitution focused more on objectivity, impartiality, and accessibility, and tended to promote desired behavior rather than avoid undesired behavior.

\begin{figure*}[h!]
  \includegraphics[width=\textwidth]{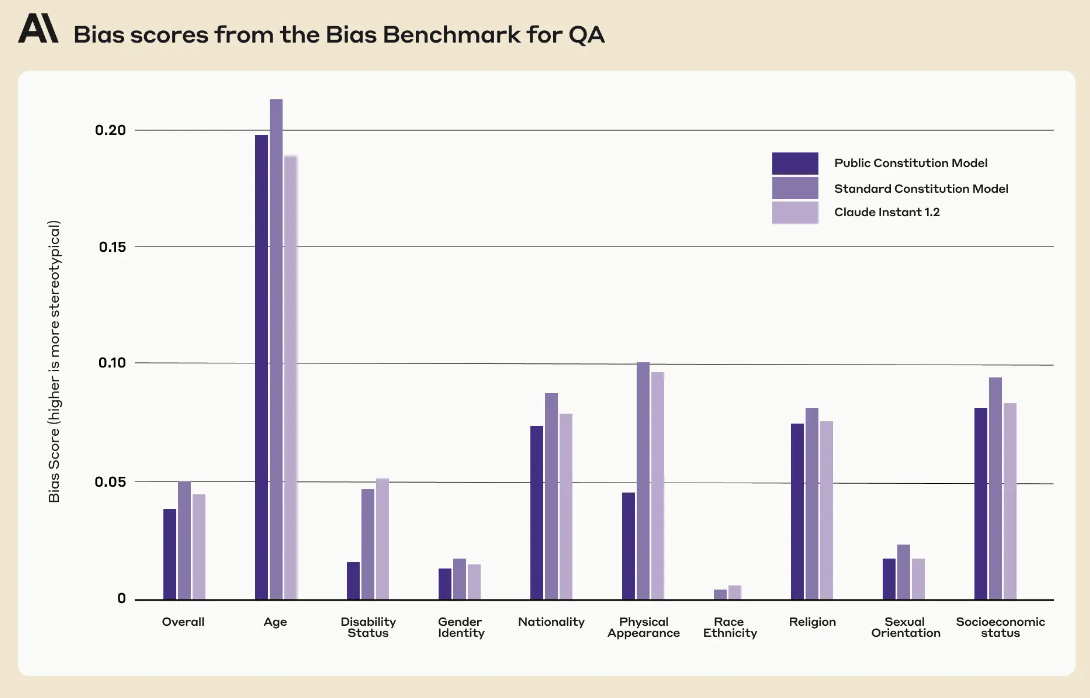}
  \caption{"BBQ\cite{parrish-etal-2022-bbq} bias scores. Higher scores indicate more negative stereotype bias (lower is better). We used the same methods, code, and controls from our previously published work. The Public model shows lower bias scores across all nine social dimensions than the Standard model, especially for Disability Status and Physical Appearance. The Public constitution places a larger emphasis on accessibility, which may explain the greater reduction in bias for Disability Status in particular."\cite{CollectiveConstitution}}
  \label{fig:bbq_bias_scores}
\end{figure*}

Anthropic trained two models using Constitutional AI: one with the publicly sourced constitution and another with their in-house constitution. The models performed similarly on language understanding and math tasks, and were perceived as equally helpful and harmless by human evaluators. However, the model trained with the public constitution showed lower bias scores(Figure:~\ref{fig:bbq_bias_scores}) across nine social dimensions, particularly in the areas of disability status and physical appearance.

\section{Threat Analysis}
\subsection{Identified Issues}

\begin{figure*}
    \centering
    \includegraphics[width=\textwidth]{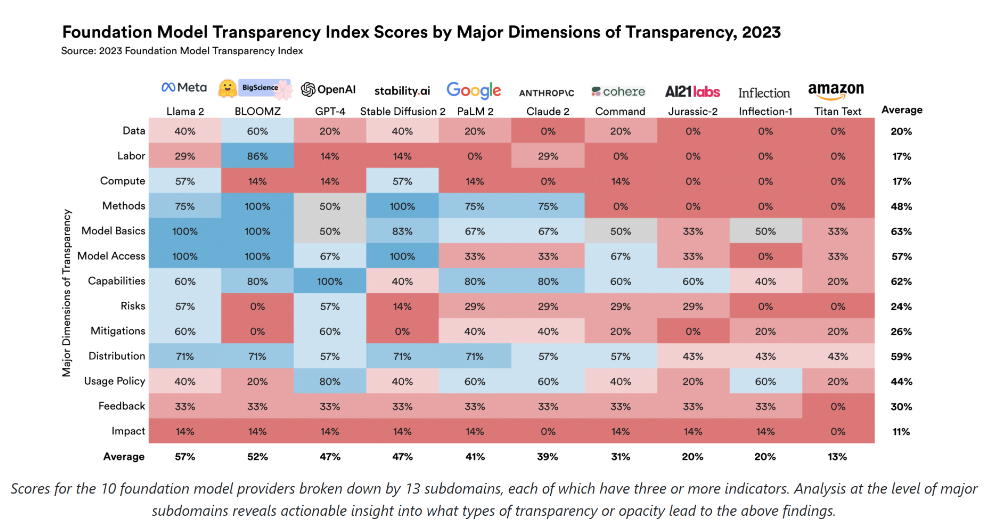}
    \caption{In depth review of Claude's feature for Foundation Model Transparency as presented in Stanford's Foundation Model Transparency Index \cite{bommasani2023foundation}.}
    \label{fig:transparency_1}
\end{figure*}

\begin{figure*}
    \centering
    \includegraphics[width=0.84\textwidth]{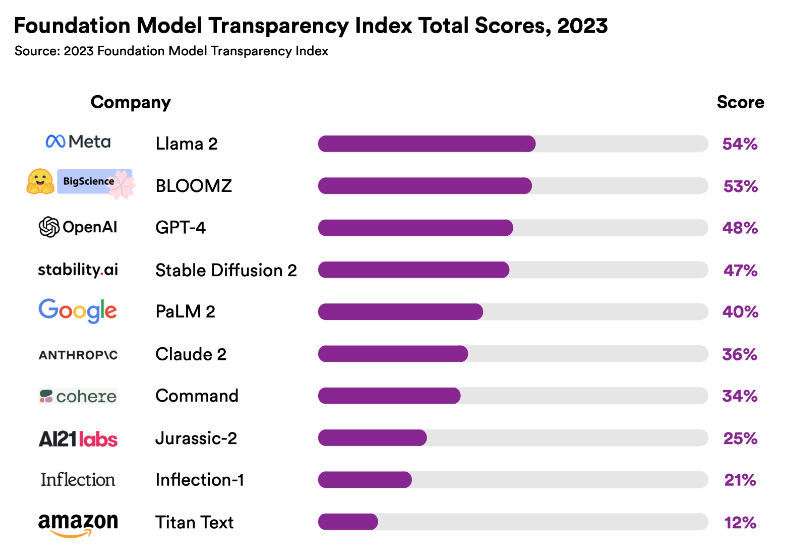}
    \caption{Results of Stanford's Foundation Model Transparency Index, places Claude really low in comparison \cite{bommasani2023foundation}.}
    \label{fig:transparency}
\end{figure*}

Through our analysis of Anthropic's Claude, we have identified several potential threats and issues that warrant attention. One significant concern is the lack of transparency in Anthropic's privacy policies, particularly regarding the collection and use of personal data for model training \cite{bommasani2023foundation}. WeThe company's policies fail to provide clear and accessible information about data handling practices, making it difficult for users to make informed decisions about their data. Anthropic automatically collects browser information, mobile network, IP address (including information about the location of the device derived from your IP address), and identifiers (including device or advertising identifiers, probabilistic identifiers, and other unique personal or online identifiers). The inadequate transparency about personal data usage in training, employing complex terminology and lacking transparency in its trust and safety review criteria, raises concerns about data security and privacy.

Another issue is the potential for hallucinations in Claude's outputs, which can lead users to believe inaccurate or misleading information. While Anthropic claims to have reduced hallucination rates compared to competitors, the lack of open-source benchmarks and validation hinders the ability to independently verify these claims. Anthropic has not released their benchmark dataset, preventing open-source comparisons. Furthermore, Anthropic's claim that Constitutional AI will employ AI itself to train out harmful model outputs is questionable, as prior research shows significant stereotype propagation in such cases \cite{arazo2020pseudo}.

Anthropic's partnerships with tech giants such as Google and Amazon raise concerns about third-party data usage and its implications for user privacy and data security. The reliance on partner policies and the lack of clear accountability mechanisms create uncertainties about data handling practices and potential risks. Anthropic claims AI Trust and Safety commitment but partners with companies that have their own data requirements. There is insufficient disclosure on Amazon and other partners training with Anthropic data. The policies lack a defined accountability structure, emphasizing responsibility without clear accountability mechanisms.

\begin{figure}[h!]
    \centering
    \includegraphics[width=0.5\textwidth]{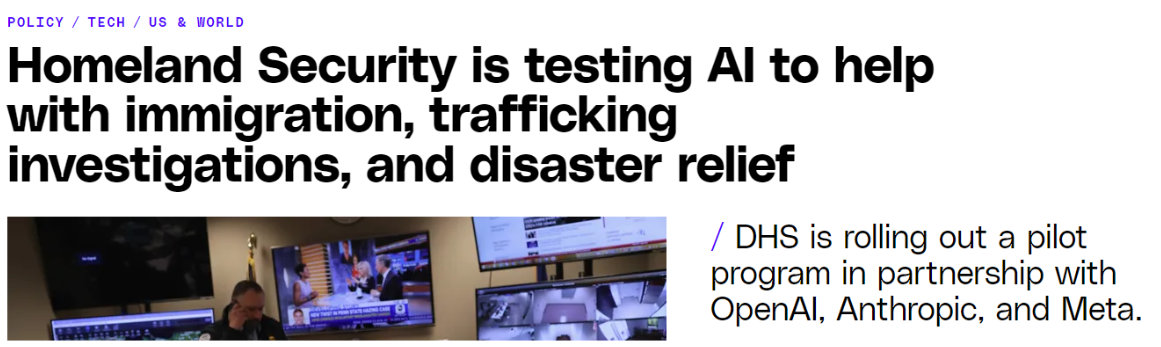}
    \caption{Department of Homeland Security working with Anthropic and other AI organizations on Pilot Programs}
    \label{fig:dhs}
\end{figure}

Potential biases and unequal benefits are another area of concern. Claude's bias benchmark, specific to Q\&A since 2022, lacks updates and may be outdated with stronger progress on red-teaming these past two years \cite{parrish-etal-2022-bbq}. Anthropic fails to disclose training data, potentially giving certain groups predisposed advantages. Biased AI can lead to unequal outcomes, particularly when implemented in government agencies like DHS and USCIS as shown in Figure~\ref{fig:dhs}, posing a high risk of discrimination \cite{DelValle2024Mar}.

Anthropic's limited engagement with relevant AI actors is another area of concern. While the company has worked with certain organizations to implement AI risk management frameworks and called for funding towards AI safety research, their engagement appears limited compared to their competitors.

Lastly, the insufficient context understanding and impact characterization across various domains raises concerns about the effectiveness of Anthropic's approach. Although the company documents and discloses their motivations and priority towards AI safety, the lack of comprehensive context understanding and impact characterization underscores the need for a more thorough approach to AI governance.

These identified issues highlight the necessity for increased transparency, accountability, and proactive measures to address potential risks associated with Anthropic's Claude. The lack of clear data usage policies, validation against open-source benchmarks, and insufficient engagement with relevant AI actors emphasizes the importance of a more comprehensive approach to AI governance. As Anthropic continues to develop and deploy its AI systems, it is crucial to address these concerns to ensure responsible and ethical AI practices.

\subsubsection{Constitutional AI}
Anthropic's Constitutional AI approach, which aims to instill fixed ethical values across all cultures, raises significant concerns. By enforcing a universal set of principles, it risks suppressing diverse perspectives, oversimplifying complex societal dynamics, and favoring certain moral frameworks while marginalizing others. The static nature of this "constitution" may struggle to adapt to evolving norms and address the nuances of translating abstract ethics into algorithms, potentially leading to unintended discriminatory consequences. Furthermore, the lack of transparency and clear public accountability mechanisms, combined with the rigidity in navigating ethical dilemmas involving conflicting principles, undermines its ability to provide nuanced ethical guidance. While well-intentioned, the Constitutional AI model's one-size-fits-all approach may inadvertently perpetuate biases encoded into its fixed framework, highlighting the need for a more dynamic, inclusive, and contextually aware ethical paradigm for responsible AI development and deployment across diverse moral landscapes.

\section{NIST Framework Analysis}

When analyzed through the lens of the NIST AI Risk Management Framework \cite{nist_ai_24}, the identified threats and issues in Anthropic's Claude can be mapped to various aspects of the framework. In terms of governance, Anthropic has defined its own AI Safety Levels and provides default opt-out options for data usage in model training. However, the company's policies lack clear accountability mechanisms, making it difficult to ensure responsible AI development and deployment.

The risk mapping and impact characterization aspects of the NIST framework reveal that Anthropic fails to appropriately disclose its objectives of AI Trust and Safety, leaving users uncertain about the risks and benefits associated with third-party software and data. While Anthropic uses a 2022 Q\&A benchmark for social bias exploration and provides model access for red-teaming and safety research, its safety-centric claims lack the proactive approach demonstrated by competitors like OpenAI.

\subsection{NIST "Govern" (Governance Analysis)}
\begin{enumerate}
    \item \textbf{Policies, processes, procedures, \& practices}: \begin{enumerate}
            \item Defined own AI Safety Levels and discloses their current models’ risks 
            \item Default opt-out for data usage in model training.
            \item Insufficient disclosure regarding the use of personal data in model training, employing complex terminology and lacking transparency in its trust and safety review criteria.
        \end{enumerate}

    \item \textbf{Accountability structure}: The policies lack a defined accountability structure, emphasizing responsibility without clear accountability. Despite Anthropic's strong recommendations in response to the NTIA's call\cite{ntia}, their policies fail to specify clear accountability mechanisms.

    \item \textbf{3rd Party Considerations}: Despite Anthropic's repeated emphasis on Trust and Safety and data protection, they often defer to their partners' policies, leaving users to decipher whether data usage is permitted. They do have a Acceptable Use Policy for API usage.

    \item \textbf{Cultural considerations \& communicated AI risks}: Presented their system prompt publicly focusing on transparency. Promoted for larger funding towards AI Safety Research. Provide special access to researchers seeking to red-team/alignment check their models.

    \item \textbf{Engagement with relevant AI Actors}: Working with NIST to implement their AI Risk Management Framework. They, also called for \$15 Million funding for NIST’s Trustworthy and Responsible AI Resource Center. Announced partnership with Google and Amazon to build for AI Safety.     
\end{enumerate}

\subsection{NIST "MAP"}
\begin{figure}
    \centering
    \includegraphics[width=0.51\textwidth]{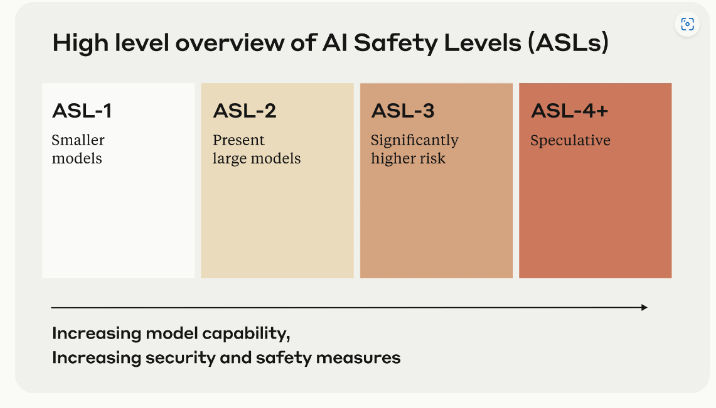}
    \caption{High level overview of AI Safety Levels defined by Anthropic\cite{APSP2024Apr}}
    \label{fig:asl}
\end{figure}

\begin{enumerate}
    \item \textbf{Context is established and understood}: Yes, Anthropic documents and discloses their motivations and priority towards AI Safety. This can be seen through their partnerships, compliance, release of own AI Safety Levels, and also bias and multilingual performance benchmarks across their models

    \item \textbf{Categorization of the AI system is performed}: Anthropic releases ASL stage for each of their models, as presented in Figure~\ref{fig:asl}. They specify tasks for biases and benchmark models for those. 

    \item \textbf{Risks and benefits are mapped to  third-party software and data}: They fail to appropriately disclose or align their objectives of AI Trust and Safety with those of their partners. For their API users, they do have a Acceptable Use Policy.

    \item \textbf{Impacts to individuals, groups, communities, organizations, and society are characterized}: Anthropic uses a 2022 Q\&A benchmark for social bias exploration and provides model access for red-teaming and safety research. However, their safety-centric claims lack OpenAI's proactive approach, which includes a curated red-teaming network actively probing for open-ended and subtle biases.
\end{enumerate}

\subsection{NIST "Manage"}

We take a look at the NIST AI Risk Management Framework's Manage function which explores Anthropic's responsibility to prioritize and respond to documented risks, plan and implement strategies to maximize benefits and minimize negative impacts, manage risks from third-party entities, and regularly monitor and document responses to identified risks \cite{nist_ai_24, NIST_AI_RMF}. By addressing these aspects, Anthropic can enhance its capacity to manage Claude's risks and ensure responsible AI development and deployment. The following points highlight the key aspects of the Manage function that Anthropic should address:

\begin{enumerate}

\item \textbf{AI risks based on impact assessments and other analytical output from the Map and Measure functions are prioritized, responded to, and managed}: Anthropic needs to prioritize and respond to the documented risks based on their potential impact, likelihood, and available resources. This includes determining whether Claude achieves its intended purpose and stated objectives, considering the risks associated with harmful usage, automation, hallucinations, biases, and weak transparency in data usage policies. Anthropic should develop, plan, and document responses to the most significant risks, which may include mitigating, transferring, sharing, avoiding, or accepting them.

\item \textbf{Strategies to maximize benefits and minimize negative impacts are planned, prepared, implemented, and documented, and informed by stakeholder input}: Anthropic should plan, prepare, implement, and document strategies to maximize the benefits and minimize the negative impacts of Claude, informed by stakeholder input. This involves considering the resources required to manage risks, along with viable alternative systems, approaches, or methods, and the related reduction in severity of impact or likelihood of each potential action. Mechanisms should be in place and applied to sustain the value of Claude and to supersede, disengage, or deactivate the system if it demonstrates performance or outcomes inconsistent with its intended use.

\item \textbf{Risks from third-party entities are managed}: Anthropic must manage risks from third-party entities, such as Google and Amazon, by regularly monitoring and applying risk controls. This is particularly important given the concerns raised about third-party data usage and its implications for user privacy and data security.

\item \textbf{Responses to identified and measured risks are documented and monitored regularly}: Anthropic should document and regularly monitor responses to identified and measured risks. This includes implementing post-deployment system monitoring plans, capturing and evaluating user and stakeholder feedback, establishing mechanisms for appeal and override, decommissioning, incident response, and change management. Measurable continuous improvement activities should be integrated into system updates and include regular stakeholder engagement.

\end{enumerate}

By addressing these aspects of the Manage function, Anthropic can enhance its capacity to manage the risks associated with Claude, allocate risk management resources based on risk measures, and ensure the responsible development and deployment of their AI system.

\subsection{NIST "MEASURE"}

The NIST AI Risk Management Framework's Measure function focuses on employing quantitative, qualitative, or mixed-method tools, techniques, and methodologies to analyze, assess, benchmark, and monitor AI risk and related impacts \cite{nist_ai_24}. Here are some points highlighting the key aspects of NIST's Measure function that Anthropic should address in relation to Claude:

\begin{enumerate}

\item \textbf{Appropriate methods and metrics are identified and applied}: Anthropic should identify and select approaches and metrics for quantitative or qualitative measurement of the most significant risks, including context-relevant measures of trustworthiness. The appropriateness of metrics and effectiveness of existing controls should be regularly assessed and updated, involving internal experts who did not serve as front-line developers for the system and/or independent assessors. Their implementation of BBQ is outdated and needs to be reconsidered \cite{parrish-etal-2022-bbq}.

\item \textbf{Systems are evaluated for trustworthy characteristics}: While Anthropic does document test sets, metrics, and details about the tools used during test, evaluation, validation, and verification (TEVV), it doesn't open-source the evaluation framework or pipeline \cite{NIST_AI_RMF}. This creates friction in academic replication tasks, who want to publicly evaluate their claims. System performance or assurance criteria should be measured qualitatively or quantitatively and demonstrated for conditions similar to deployment setting(s). Claude should be evaluated regularly for safety, computational bias, resilience, security, privacy risk, and environmental impact. The AI model should be explained, validated, and documented, and its output should be interpreted within its context to inform responsible use and governance.

\item \textbf{Mechanisms for tracking identified risks over time are in place}: Anthropic should have approaches, personnel, and documentation in place to regularly identify and track existing and emergent risks based on factors such as intended and actual performance in deployed contexts. Risk tracking approaches should be considered for settings where risks are difficult to assess using currently available measurement techniques or are not yet available. They do not have a Red Teaming Network (internal exists) like OpenAI nor do they have a Bug Bountry program yet, which could both be used as a crowd-sourced risk tracker.

\item \textbf{Feedback about efficacy of measurement is gathered and assessed}: Measurement approaches for identifying risks should be connected to deployment context(s) and informed through consultation with domain experts and other end users. Measurement results regarding system trustworthiness in deployment context(s) should be informed by domain expert and other stakeholder feedback to validate whether Claude is performing consistently as intended.

\end{enumerate}

By addressing these aspects of the Measure function, Anthropic can enhance their capacity to comprehensively evaluate Claude's trustworthiness, identify and track existing and emergent risks, and verify the efficacy of metrics.

\section{EU AI Act Analysis}
Under the EU AI Act \cite{EUAI2024Apr}, the identified threats and issues in Anthropic's Claude can be categorized based on their risk levels. The risks associated with harmful usage and automation are also significant:

\begin{enumerate}
    \item Automation of AI fine-tuning can be considered limited risk, as previous research has demonstrated it to propagate pre-learnt biases. However, if this model is auto-deployed without validation, it would become high risk.
    \item Hallucinations in outputs can cause users to mistakenly believe something, directly impacting individuals and their perception of reality, posing a limited risk.
    \item Lack of transparency can affect user trust and lead to privacy issues. If the data is used for training other models, it can be considered high risk because AI models can memorize information.
    \item AI being used with the intent to cause harm, such as violating human rights, poses an unacceptable risk.
\end{enumerate}

As discussed, under the EU AI Act, the use of AI for harmful content removal, as proposed in Anthropic's Constitutional AI framework, would likely be classified as a high-risk AI system. This is because the automated removal of content can have significant impacts on individuals' rights to freedom of expression and access to information. The lack of transparency and potential for biases in the AI system used for content moderation further exacerbates the risks associated with this application. Similarly, the use of Claude in government agencies like DHS and USCIS would also fall under the high-risk category due to the potential for discriminatory outcomes and unequal treatment. While this maybe within the United States, we still evaluate it using the EU AI Framework, which emphasizes the importance of ensuring that AI systems used in the public sector are transparent, accountable, and free from biases that could lead to discrimination.

Anthropic's weak transparency in data usage policies and the potential for data used in training to be memorized and reproduced by AI models raise concerns under the EU AI Act's requirements for data governance and privacy protection. The Act requires AI system providers to ensure appropriate data management practices, including data minimization, data quality, and data protection safeguards. To comply with the EU AI Act, Anthropic would need to address these identified risks by implementing robust risk management processes, ensuring transparency in their AI systems' development and deployment, and establishing clear accountability mechanisms.


\section{Proposed Mitigations \& Resolution Strategies}
To address the identified threats and issues in Anthropic's Claude, we propose the following mitigation strategies:

\subsection{Enhance transparency in privacy policies}
Anthropic should prioritize adopting transparent privacy practices that comprehensively detail the risks associated with artificial intelligence systems, as outlined in the NIST AI Framework \cite{nist_ai_24}. Additionally, they should minimize data retention periods and implement a default opt-out option, empowering users with greater control over their personal information. To further simplify information access and boost user engagement, organizations should streamline navigation complexity and provide concise, easily understandable summaries of their privacy practices. This will empower users to make informed decisions about their data and increase trust in Anthropic's AI systems.

\subsubsection{Criteria \& Metrics}

The evaluation of efforts to improve the transparency and accessibility of privacy policies should be guided by well-defined criteria and metrics. These include:
\begin{enumerate}
    \item \textbf{Accessibility}: Measured by the average number of clicks required for users to access the privacy practices. A lower number of clicks indicates higher accessibility, enabling users to obtain privacy policy information more conveniently.

    \item \textbf{Time}: The duration spent by users locating specific details within the privacy policy. This metric assesses the ease with which users can quickly find the required information within the policy. A shorter duration reflects better organization and navigation of the privacy policy.

    \item \textbf{Comprehension}: The extent to which users can understand the content of the privacy policies without relying on external references. This metric evaluates the clarity and readability of the policies; the clearer the language, the easier it is for users to comprehend without requiring external explanations.
\end{enumerate}

\subsubsection{Data Sources / Test}

To collect data for these metrics, two primary methods can be employed:

Comprehension Surveys: Designing questionnaires that present users with the privacy policy content and assess their understanding through questions. The survey results can provide valuable insights into the comprehension metric.

Benchmarking: Comparing the organization's metrics against industry standards or best practices. By benchmarking their accessibility, time, and comprehension metrics against established norms or leading examples, organizations can identify areas for improvement and gauge their performance relative to peers or competitors.

Utilizing these data sources and testing methods, organizations can gather valuable data and insights to evaluate the effectiveness of their efforts in improving the transparency and accessibility of privacy policies. This information will guide further improvements and help organizations prioritize areas that require the most attention and resources.

\subsubsection{Practical Considerations}

From a practical standpoint, organizations should focus on enhancing user experience through thoughtful design choices and effective summarization techniques. Simplifying the interface and reducing navigation complexity can expedite information access while offering clear and concise summaries of privacy practices can significantly improve user comprehension and engagement with these policies.

\subsection{Establish rigorous benchmarks for hallucination and bias} 
To ensure transparency and facilitate rigorous public scrutiny of potential hallucinations and biases in Anthropic's AI models, it is imperative to conduct comprehensive benchmarking exercises. 

\subsubsection{Criteria \& Metrics}
These benchmarks should aim to measure the extent of hallucinations, which can be quantified through metrics such as Q2 and factual-grounding BLEU scores. Additionally, they should evaluate various forms of bias, including statistical parity, group diversity, equalized odds, and even open-ended opinions annotated by subject matter experts. This will help identify and mitigate potential risks associated with inaccurate or biased outputs.

\subsubsection{Data Sources / Test}
The data sources and tests employed for these benchmarking efforts should be diverse and comprehensive. For hallucination evaluation, datasets such as HaluEval \cite{Li2023Dec} and the forthcoming HaluEval-Wild \cite{Zhu2024Mar} can provide valuable insights. Bias assessment can leverage resources like R-Judge \cite{Yuan2024Jan}, CBBQ \cite{Huang2023Jun}, Winoqueer \cite{Felkner2023Jun} and KorNAT \cite{Lee2024Feb}, which cover a wide range of bias types and demographic factors.

\subsubsection{Practical Considerations}
From a practical standpoint, it is crucial to ensure that these benchmarks are not inadvertently used for pre-training or fine-tuning the AI models themselves, as this could introduce biases or undermine the integrity of the evaluation process. Additionally, creating private leaderboards for these benchmarks can help maintain their integrity and prevent potential gaming or exploitation.

\subsection{Develop a comprehensive remediation process} 
Anthropic should implement a robust process for handling user requests for data deletion and ensuring the unlearning of data by AI models. Clear mechanisms should be established for users to initiate data deletion requests, and the company should provide detailed guidance and support throughout the process. Rigorous testing should be conducted to verify the effectiveness of data removal and model unlearning.

\subsubsection{Criteria \& Metrics}
This remediation process should prioritize clarity in the request initiation stage, empowering users with a straightforward understanding of how to initiate data deletion requests. Furthermore, it must incorporate verifiable metrics to assess the efficacy of model unlearning processes, ensuring that user data is thoroughly expunged from the models upon request.

\subsubsection{Data Sources / Test}
User feedback can provide invaluable insights into the clarity and user-friendliness of the data deletion process, highlighting areas that may require improvement. Additionally, dedicated unlearning tests must be conducted to verify the complete removal of user data from the models, validating the integrity and functionality of the unlearning mechanisms.

\subsubsection{Practical Considerations}
From a practical standpoint, several key considerations must be addressed. Firstly, the entire process of data deletion and unlearning should be transparent, with clear and well-documented steps outlined for users. Secondly, comprehensive guidance and user support should be provided throughout the process, ensuring that users are adequately informed and assisted at every stage. Finally, organizations must invest in enhancing their systems to support efficient and timely data deletion and unlearning, prioritizing the swift and effective handling of user requests.

By implementing these mitigation strategies, Anthropic can demonstrate its commitment to responsible AI development and deployment, enhance user trust, and reduce the risks associated with its Claude model. Regular monitoring and continuous improvement of these measures will be essential to keep pace with the evolving AI governance landscape and ensure ongoing accountability.

\section{Discussion}
The proposed mitigation strategies for Anthropic's Claude have significant implications for the broader AI governance landscape and the social impact of AI systems. By enhancing transparency in privacy policies, Anthropic can set a positive example for other AI companies, encouraging a more open and accountable approach to data handling and user privacy. This increased transparency will empower users to make informed decisions about their data and foster trust in AI systems.

Establishing rigorous benchmarks for hallucination and bias will contribute to the development of more reliable and unbiased AI models. By publicly releasing datasets and results, Anthropic can promote collaboration and knowledge sharing within the AI community, driving collective efforts towards mitigating the risks associated with inaccurate or biased outputs. This transparency will also enable independent verification and accountability, ensuring that AI systems are subject to rigorous scrutiny.

Implementing a comprehensive remediation process for data deletion and model unlearning will address concerns about data privacy and the potential misuse of personal information. By providing users with clear mechanisms to control their data and ensuring the effectiveness of data removal and model unlearning, Anthropic can demonstrate its commitment to user privacy and build trust in its AI systems.

The adoption of these mitigation strategies by Anthropic and other AI companies will contribute to the development of a more responsible and trustworthy AI ecosystem. As AI systems become increasingly integrated into various aspects of society, ensuring their alignment with ethical principles and societal values becomes paramount. By prioritizing transparency, accountability, and user privacy, AI companies can foster public trust and support the responsible deployment of AI technologies for the benefit of society.

\section{Conclusion}
In conclusion, this paper has examined the AI governance landscape, focusing on Anthropic's Claude as a case study~\cite{Parrish2022May}. Through the lens of the NIST AI Risk Management Framework and the EU AI Act, we have identified potential threats and issues in Claude, including the lack of transparency in privacy policies, the potential for hallucinations and biases in outputs, and concerns about third-party data usage\cite{Yao2023May}. To address these challenges, we have proposed mitigation strategies that emphasize transparency, rigorous benchmarking, and comprehensive data handling processes. By adopting these measures, Anthropic can demonstrate its commitment to responsible AI development and deployment, enhance user trust, and contribute to the broader efforts in AI governance.

The evolution of AI governance will require ongoing collaboration, adaptation, and learning from the successes and challenges of parallel domains such as privacy regulations. As AI systems become more sophisticated and integrated into society, ensuring their alignment with ethical principles and societal values will be critical. By prioritizing accountability, transparency, and user privacy, AI companies can foster public trust and support the responsible advancement of AI technologies for the benefit of society.

\section{Limitations \& Ethical Considerations}
While this paper provides valuable insights into AI governance and accountability, it is important to acknowledge its limitations. The analysis focuses primarily on Anthropic's Claude and may not fully capture the diverse range of AI systems and their unique governance challenges. Additionally, the proposed mitigation strategies, while promising, require further validation and real-world implementation to assess their effectiveness and potential unintended consequences.

Ethical considerations are paramount in the development and deployment of AI systems. As AI technologies become more powerful and influential, it is crucial to ensure that they are designed and used in a manner that respects human rights, promotes fairness, and avoids harmful biases. AI companies must prioritize ethical principles throughout the AI lifecycle, from data collection and model training to deployment and monitoring.

Ongoing research, collaboration, and stakeholder engagement will be essential to address the ethical implications of AI and develop robust governance frameworks that keep pace with the rapid advancements in AI technologies. By proactively addressing ethical considerations and prioritizing accountability, transparency, and user privacy, we can work towards a future where AI systems are trusted, beneficial, and aligned with societal values.

\section{Acknowledgments}
I would like to express my sincere gratitude to Professor Norman Sadeh for his invaluable guidance and insights throughout the AI Governance course at Carnegie Mellon University. We are truly thankful for the opportunity to learn about the field of AI governance.

\bibliographystyle{IEEEtran} 
\bibliography{reference}

\end{document}